\documentclass{elsart}
\usepackage{amsfonts}
\usepackage{amsmath}
\usepackage{graphicx}
\usepackage{amssymb}

\begin{document}

\begin{frontmatter}
\title{Vacuum polarization and plasma oscillations}
\author{R. Ruffini}
\ead{Ruffini@icra.it}
\author{, G.V. Vereshchagin, S.-S. Xue}
\address{ICRANet, P.le della Repubblica 10, 65100 Pescara, Italy, \\
ICRA and University of Rome "Sapienza", Physics Department, \\
P.le A. Moro 5, 00185 Rome, Italy.}

\begin{abstract}
We evidence the existence of plasma oscillations of electrons-positron pairs created by the vacuum polarization in an uniform electric field with $E\lesssim E_{c}$. Our general treatment, encompassing also the traditional, well studied case of $E>E_{c}$, shows the existence in both cases of a maximum Lorentz factor acquired by electrons and positrons and allows determination of the a maximal length of oscillation. We quantitatively estimate how plasma oscillations reduce the rate of pair creation and increase the time scale of the pair production. These results are particularly relevant in view of the experimental progress in approaching the field strengths $E\lesssim E_{c}$.
\end{abstract}



\end{frontmatter}

The plasma oscillation phenomenon of the electron-positron pairs created by
vacuum polarization for $E>E_{c}\equiv m^{2}c^{3}/(e\hbar )$ ($\hbar $ is
the Planck's constant, $c$ is the speed of light, $m$ and $e$ are the
electron mass and the absolute value of electron charge respectively)
represents one of the most popular topics in relativistic field theory
today. In particular the Vlasov-Boltzmann equation has been used within QCD
in \cite{osc},\cite{osc1}, and with the semiclassical field equations in 
\cite{osc2} within QED. This approach was shown in \cite{osc2} to be in
excellent agreement with quantum field theory calculations \cite{QFT}.
Applications of these studies range from heavy ion collisions \cite{ion}-%
\cite{ion2} to lasers \cite{Laser}.

We have introduced collisional terms in the Vlasov-Boltzmann equation for
such a system in \cite{Astro}. Our results have been considered of interest
in the studies of pair production in free electron lasers \cite{Ringwald},%
\cite{Laser1}, in optical lasers \cite{Laser2}, of millicharged fermions in
extensions of the standard model of particle physics \cite{Gies},
electromagnetic wave propagation in a plasma \cite{Bulanov}, as well in
astrophysics \cite{dya2}.

In this Letter we explore the case of undercritical electric field which has
not yet been studied in the literature. It is usually expected that for $%
E<E_{c}$ back reaction of the created electrons and positrons on the
external electric field can be neglected and electrons and positrons would
move as test particles along electric lines of force. Here we show that this
is not the case in a uniform unbounded field. This work is urgent since the
first observation of oscillations effects should be first detectable in
experiments for the regime $E<E_{c}$, in view of the rapid developments in
experimental techniques, see e.g. \cite{Tajima},\cite{Bulanov2}.

We introduce an approach based on continuity, energy-momentum conservation
and Maxwell equations in order to account for the back reaction of the
created pairs. By this treatment we can analyse the new case of undecritical
field, $E<E_{c}$, and recover the old results for overcritical field, $%
E>E_{c}$. In particular, we are focusing on the range $0.15E_{c}<E<10E_{c}$.

It is generally assumed that electrons and positrons are created at rest in
pairs, due to vacuum polarization in uniform electric field with strength $E$
\cite{S}-\cite{book2}, with the average rate per unit volume and per unit
time\footnote{%
We use in the following the system of units where $\hbar=c=1$, $e=\sqrt{%
\alpha}\approx\sqrt{1/137}$, $\alpha$ being the fine structure constant.} 
\begin{equation}
S\equiv\frac{dN}{dVdt}=\frac{m^{4}}{4\pi^{3}}\left( \frac{E}{E_{c}}\right)
^{2}\exp\left( -\pi\frac{E_{c}}{E}\right) .   \label{rate}
\end{equation}
This formula is derived for uniform constant in time electric field.
However, it still can be used for slowly time-varying electric field
provided the inverse adiabaticity parameter \cite{book1}-\cite{Popov} is
much larger than one,%
\begin{equation}
\eta=\frac{m}{\omega}\frac{E_{peak}}{E_{c}}=\tilde{T}\tilde{E}_{peak}\gg1, 
\label{eta}
\end{equation}
where $\omega$ is the frequency of oscillations, $\tilde{T}=m/\omega$ is
dimensionless period of oscillations. Equation (\ref{eta}) implies that time
variation of the electric field is much slower than the rate of pair
production. In two specific cases considered in this paper, $E=10E_{c}$ and $%
E=0.15E_{c}$ we find for the first oscillation $\eta=334$ and $\eta
=3.1\times10^{6}$ respectively. This demonstrates applicability of the
formula (\ref{rate}) in our case.

From the continuity, energy-momentum conservation and Maxwell equations
written for electrons, positrons and electromagnetic field we have 
\begin{align}
\frac{\partial\left( \bar{n}U^{\mu}\right) }{\partial x^{\mu}} & =S,
\label{cont} \\
\frac{\partial T^{\mu\nu}}{\partial x^{\nu}} & =-F^{\mu\nu}J_{\nu },
\label{em} \\
\frac{\partial F^{\mu\nu}}{\partial x^{\nu}} & =-4\pi J^{\mu},   \label{me}
\end{align}
where $\bar{n}$ is the comoving number density of electrons, $T^{\mu\nu}$ is
energy-momentum tensor of electrons and positrons 
\begin{equation}
T^{\mu\nu}=m\bar{n}\left(
U_{(+)}^{\mu}U_{(+)}^{\nu}+U_{(-)}^{\mu}U_{(-)}^{\nu}\right) , 
\label{emten}
\end{equation}
$F^{\mu\nu}$ is electromagnetic field tensor, $J^{\mu}$ is the total
four-current density, $U^{\mu}$ is four velocity respectively of positrons
and electrons%
\begin{equation}
U_{(+)}^{\mu}=U^{\mu}=\gamma\left( 1,v,0,0\right) ,\qquad U_{(-)}^{\mu
}=\gamma\left( 1,-v,0,0\right) ,
\end{equation}
$v$ is the average velocity of electrons, $\gamma=\left(1-v^{2}\right)^{-1/2}
$ is relativistic Lorentz factor. Electrons and positrons move along the
electric field lines in opposite directions.

We choose a coordinate frame where pairs are created at rest. Electric field
in this frame is directed along $x$-axis and introduce coordinate number
density $n=\bar{n}\gamma$. In spatially homogeneous case from (\ref{cont})
we have%
\begin{equation}
\dot{n}=S.
\end{equation}
With our definitions (\ref{emten}) from (\ref{em}) and equation of motion
for positrons and electrons%
\begin{equation}
m\frac{\partial U_{(\pm)}^{\mu}}{\partial x^{\nu}}=\mp eF_{\nu}^{\mu},
\end{equation}
we find 
\begin{equation}
\frac{\partial T^{\mu\nu}}{\partial x^{\nu}}=-e\bar{n}\left( U_{(+)}^{\nu
}-U_{(-)}^{\nu}\right) F_{\nu}^{\mu}+mS\left( U_{(+)}^{\mu}+U_{(-)}^{\mu
}\right) =-F_{\nu}^{\mu}J^{\nu},
\end{equation}
where the total current density is the sum of conducting $J_{cond}^{\mu}$
and polarization $J_{pol}^{\mu}$ currents \cite{ion1} densities%
\begin{align}
\qquad J^{\mu} & =J_{cond}^{\mu}+J_{pol}^{\mu}, \\
J_{cond}^{\mu} & =e\bar{n}\left( U_{(+)}^{\mu}-U_{(-)}^{\mu}\right) , \\
J_{pol}^{\mu} & =\frac{2mS}{E}\gamma\left( 0,1,0,0\right) .
\end{align}

Energy-momentum tensor in (\ref{em}) and electromagnetic field tensor in (%
\ref{me}) change for two reasons: 1)\ electrons and positrons acceleration
in the electric field, given by the term $J_{cond}^{\mu}$, 2)\ particle
creation, described by the term $J_{pol}^{\mu}$. Equation (\ref{cont}) is
satisfied separately for electrons and positrons.

Defining energy density of positrons%
\begin{equation}
\rho=\frac{1}{2}T^{00}=mn\gamma,
\end{equation}
we find from (\ref{em}) 
\begin{equation}
\dot{\rho}=envE+ m\gamma S.
\end{equation}
Due to homogeneity of the electric field and plasma, electrons and positrons
have the same energy and absolute value of the momentum density $p$, but
their momenta have opposite directions. Our definitions also imply for
velocity and momentum densities of electrons and positrons%
\begin{equation}
v=\frac{p}{\rho},   \label{veleq}
\end{equation}
and%
\begin{equation}
\rho^{2}=p^{2}+m^{2}n^{2},   \label{rhopn}
\end{equation}
which is just relativistic relation between the energy, momentum and mass
densities of particles.

Gathering together the above equations we then have the following equations%
\begin{align}
\dot{n}& =S,  \label{ndot} \\
\dot{\rho}& =E\left( env+\frac{m\gamma S}{E}\right) ,  \label{rhodot} \\
\dot{p}& =enE+mv\gamma S,  \label{pdot} \\
\dot{E}& =-8\pi \left( env+\frac{m\gamma S}{E}\right) .  \label{Edot}
\end{align}%
From (\ref{rhodot}) and (\ref{Edot}) we obtain the energy conservation
equation%
\begin{equation}
\frac{E_{0}^{2}-E^{2}}{8\pi }+2\rho =0,  \label{energy}
\end{equation}%
where $E_{0}$ is the constant of integration, so the particle energy density
vanishes for initial value of the electric field, $E_{0}$.

These equations give also the maximum number of the pair density
asymptotically attainable consistently with the above rate equation and
energy conservation%
\begin{equation}
n_{0}=\frac{E_{0}^{2}}{8\pi m}.   \label{n0}
\end{equation}

For simplicity we introduce dimensionless variables $n=m^{3}\tilde{n}$, $%
\rho=m^{4}\tilde{\rho}$, $p=m^{4}\tilde{p}$, $E=E_{c}\tilde{E}$, and $%
t=m^{-1}\tilde{t}$. With these variables our system of equations (\ref{ndot}%
)-(\ref{Edot}) takes the form 
\begin{align}
\frac{d\tilde{n}}{d\tilde{t}} & =\tilde{S},  \notag \\
\frac{d\tilde{\rho}}{d\tilde{t}} & =\tilde{n}\tilde{E}\tilde{v}+\tilde{\gamma%
}\tilde{S},  \label{numsys} \\
\frac{d\tilde{p}}{d\tilde{t}} & =\tilde{n}\tilde{E}+\tilde{\gamma}\tilde {v}%
\tilde{S},  \notag \\
\frac{d\tilde{E}}{d\tilde{t}} & =-8\pi\alpha\left( \tilde{n}\tilde{v}+\frac{%
\tilde{\gamma}\tilde{S}}{\tilde{E}}\right) ,  \notag
\end{align}
where $\tilde{S}=\frac{1}{4\pi^{3}}\tilde{E}^{2}\exp\left( -\frac{\pi}{%
\tilde{E}}\right) $, $\tilde{v}=\frac{\tilde{p}}{\tilde{\rho}}$ and $\tilde{%
\gamma}=\left( 1-\tilde{v}^{2}\right) ^{-1/2}$, $\alpha=e^{2}/(\hbar c)$ as
before.

We solve numerically the system of equations (\ref{numsys}) with the initial
conditions $n(0)=\rho(0)=v(0)=0$, and the electric field $E(0)=E_{0}$.

\begin{figure}[ptb]
\begin{center}
\includegraphics[width=5in]{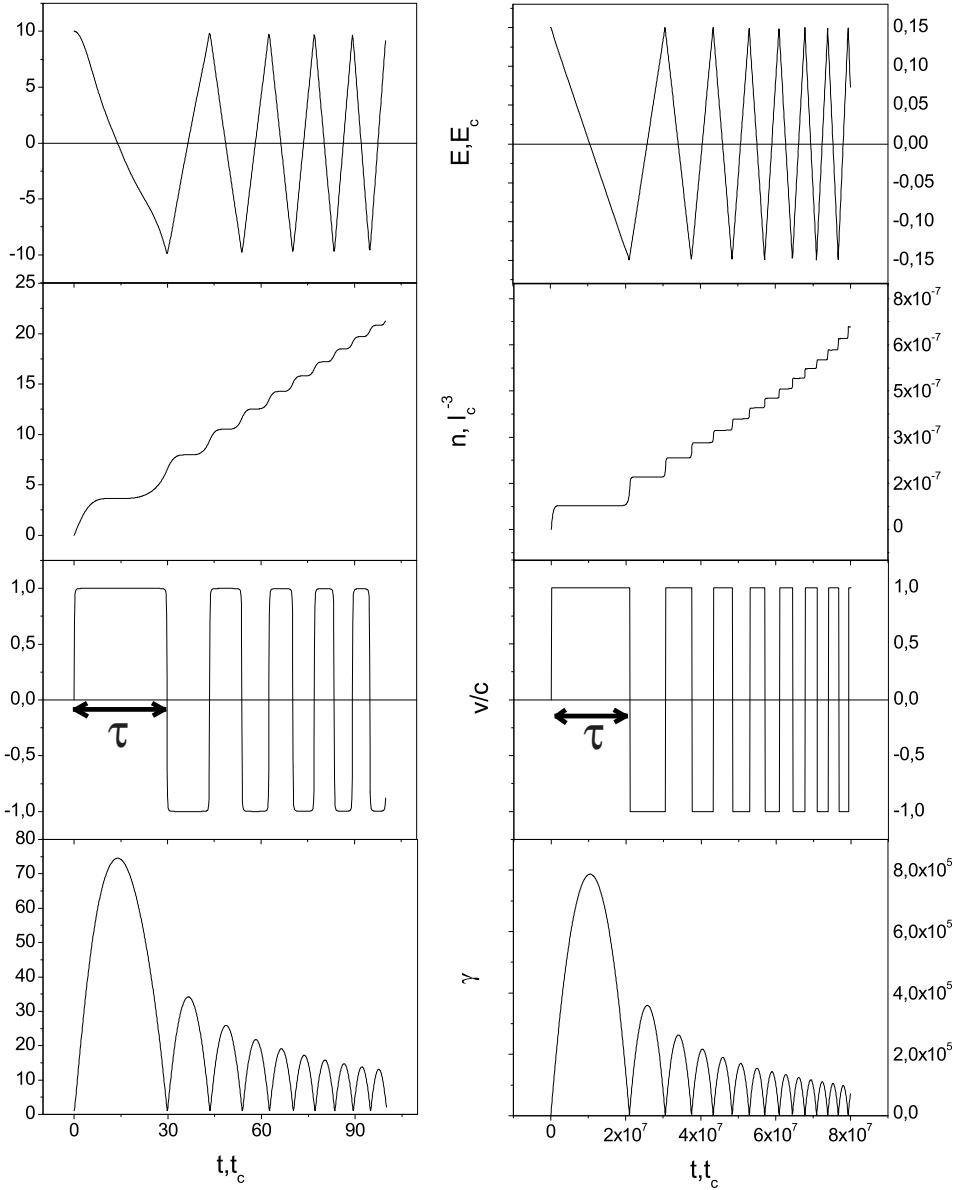}
\end{center}
\caption{Electric field strength, number density of electrons, their
velocity and Lorentz gamma factor depending on time with $E_{0}=10E_{c}$
(left column) and $E_{0}=0.15E_{c}$ (right column). Electric field, number
density and velocity of positron are measured respectively in terms of the
critical field $E_{c}$, Compton volume $l_{c}^{3}=\left( \frac{\hbar}{mc}%
\right) ^{3}$, and the speed of light $c$. We define the length of
oscillation as $D=c\protect\tau$, where $\protect\tau$ is the time needed
for the first half-oscillation, shown above.}
\label{fig1}
\end{figure}

In fig. \ref{fig1} we provide diagrams for electric field strength, number
density, velocity and Lorentz gamma factor of electrons as functions of
time, for initial values of the electric field $E_{0}=10E_{c}$ (left column)
and $E_{0}=0.15E_{c}$ (right column). Slowly decaying plasma oscillations
develop in both cases. We estimated the half-life of oscillations to be $%
10^{3}t_{c}$ for $E_{0}=10E_{c}$ and $10^{5}t_{c}$ for $E_{0}=0.8E_{c}$
respectively. The period of the fist oscillation is $50t_{c}$ and $%
3\times10^{7}t_{c}$, the Lorentz factor of electrons and positrons in the
first oscillation equals $75$ and $3\times10^{5}$ respectively for $%
E_{0}=10E_{c}$ and $E_{0}=0.15E_{c}$. Therefore, in contrast to the case $%
E>E_{c}$, for $E<E_{c}$ plasma oscillations develop on a much longer
timescale, electrons and positrons reach extremely relativistic velocities.

\begin{figure}[th]
\begin{center}
\includegraphics[width=5.5in]{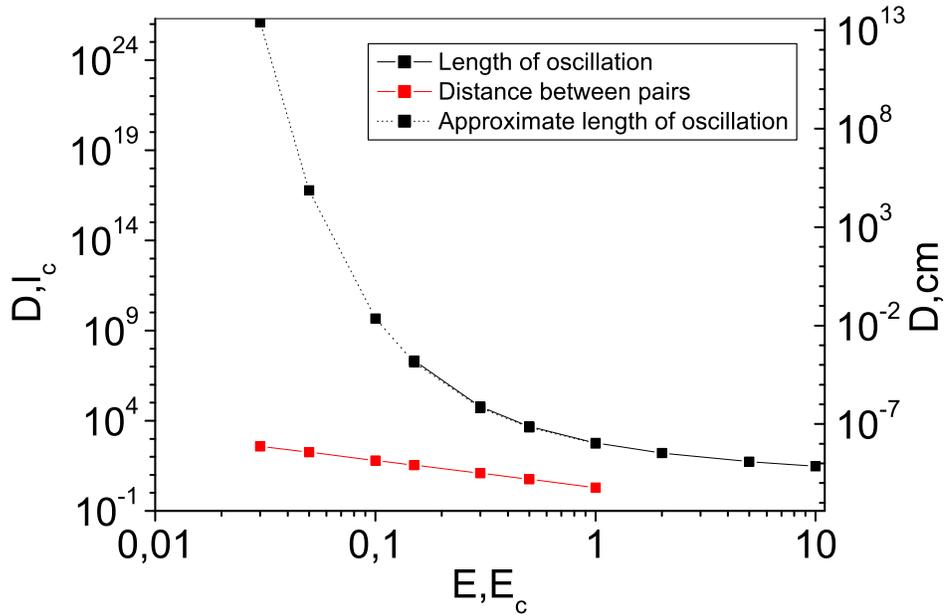}
\end{center}
\caption{Maximum length of oscillations (black curves) together with the
distance between electron and positron in a pair (red curve) computed from (%
\protect\ref{D}), depending on initial value of electric field strength. The
solid black curve is obtained from solutions of exact equations (\protect\ref%
{numsys}), while the dotted black curve corresponds to solutions of
approximate equation (\protect\ref{eeq}).}
\label{fig2}
\end{figure}

In fig. \ref{fig2} the characteristic length of oscillations is shown
together with the distance between the pairs at the moment of their
creation. For constant electric field the formation length for the
electron-positron pairs, or the quantum tunnening length, is not simply $%
m/(eE)$, as expected from a semi-classical approximation, but \cite{Pair},%
\cite{Pair1} 
\begin{equation}
D^{\ast}=\frac{m}{eE}\left( \frac{E_{c}}{E}\right) ^{1/2}.   \label{D}
\end{equation}

Thus, given initial electric field strength we define two characteristic
distances: $D^{\ast}$, the distance between created pairs, above which pair
creation is possible, and the length of oscillations, $D=c\tau$, above which
plasma oscillations occur in a uniform electric field. The length of
oscillations is the maximal distance between two turning points in the
motion of electrons and positrons (see fig. \ref{fig2}). From fig. \ref{fig2}
it is clear that $D\gg D^{\ast}$. In the oscillation phenomena the larger
electric field is, the larger becomes the density of pairs and therefore the
back reaction, or the screening effect, is stronger. Thus the period of
oscillations becomes shorter. Note that the frequency of oscillation is not
equal to the plasma frequency, so it cannot be used as the measure of the
latter. Notice that for $E\ll E_{c}$ the length of oscillations becomes
macroscopically large.
\begin{figure}[tbh]
\begin{center}
\includegraphics[width=5.5in]{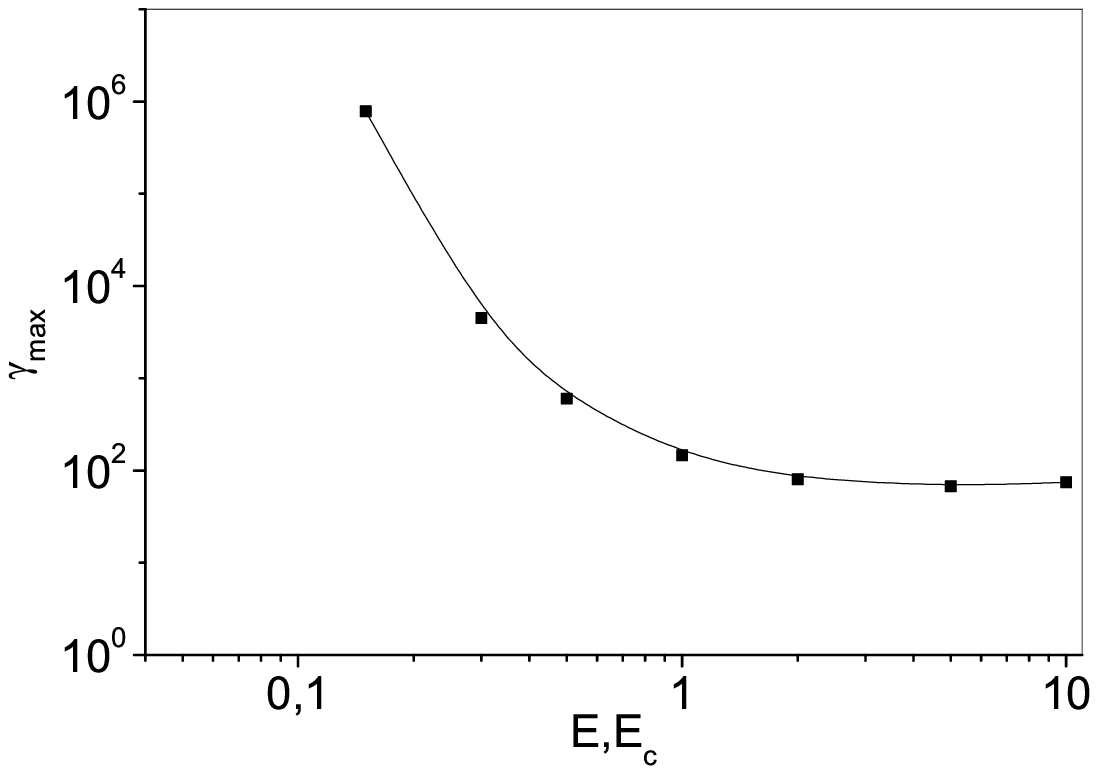}
\end{center}
\caption{Maximum Lorentz gamma factor $\protect\gamma$ reached at the first
oscillation depending on initial value of the electric field strength.}
\label{fig3}
\end{figure}

At fig. \ref{fig3} maximum Lorentz gamma factor in the first oscillation is
presented depending on initial value of the electric field. Since in the
successive oscillations the maximal value of the Lorentz factor is
monotonically decreasing (see fig. \ref{fig1}) we conclude that for every
initial value of the electric field there exists a maximum Lorentz factor
attainable by the electrons and positrons in the plasma. It is interesting
to stress the dependence of the Lorentz factor on initial electric field
strength. The kinetic energy contribution becomes overwhelming in the $%
E<E_{c}$ case. On the contrary, in the case $E>E_{c}$ the electromagnetic
energy of the field goes mainly into the rest mass energy of the pairs.

This diagram clearly shows that never in this process the test particle
approximation for the electrons and positrons motion in uniform electric
field can be applied. Without considering back-reaction on the initial
field, electrons and positrons moving in a uniform electric field would
experience constant acceleration reaching $v\sim c$ for $E=E_{c}$ on the
timescale $t_{c}$ and keep that speed thereafter. Therefore, the back
reaction effects in a uniform field are essential both in the case of $%
E>E_{c}$ and $E<E_{c}$.

\begin{figure}[tph]
\centering
\includegraphics[width=5in]{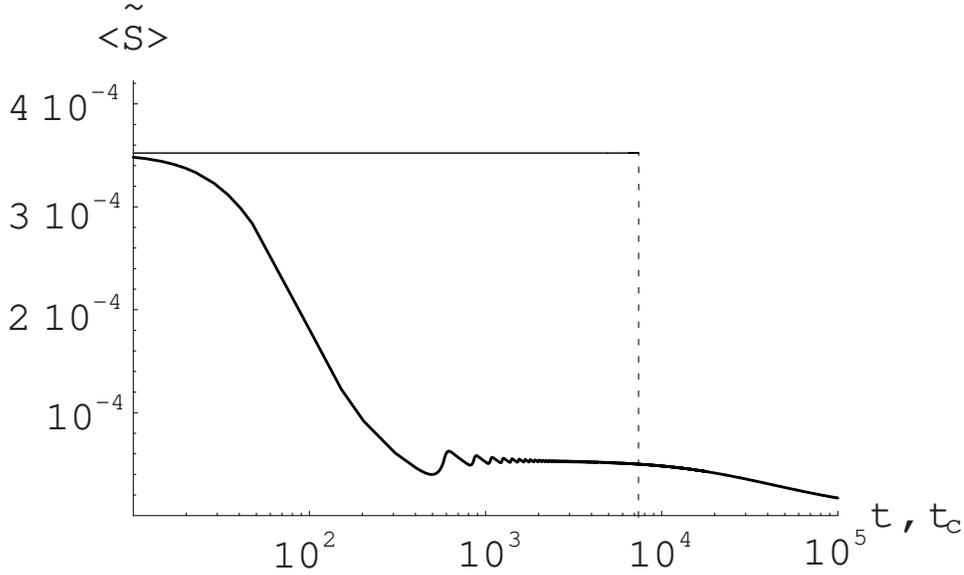}
\caption{The average rate of pair production $n/t$ is shown as function of
time (thick curve), comparing to its initial value $S(E_{0})$ (thin line)
for $E_{0}=E_{c}$. The dashed line marks the time when the energy of
electric field would have exhaused if the rate kept constant.}
\label{fig4}
\end{figure}

We compare the average rate of pair creation for two cases:\ when the
electric field value is constant in time (an external energy source keeps
the field unchanged) and when it is self-regulated by equations (\ref{numsys}%
). The result is represented in fig. \ref{fig4}. It is clear from fig. \ref%
{fig4} that when the back reaction effects are taken into account, the
effective rate of the pair production is smaller than the corresponding rate
(\ref{rate}) in a uniform field $E_{0}$. At the same time, discharge of the
field takes much longer time. To quantify this effect we compute the
efficiency of the pair production defined as $\epsilon =n(t_{S})/n_{0}$
where $t_{S}$ is the time when pair creation with the constant rate $S(E_{0})
$\ would stop, and $n_{0}$ is defined above, see (\ref{n0}). For $E_{0}=E_{c}
$ we find $\epsilon =14$\%, while for $E_{0}=0.3E_{c}$ we have $\epsilon =1$%
\%.

It is clear from the structure of the above equations that for $E<E_{c}$ the
number of pairs is small, electrons and positrons are accelerated in
electric field and the conducting current dominates. Assuming electric field
to be weak we neglect polarization current in energy conservation (\ref%
{rhodot}) and in Maxwell equation (\ref{Edot}). This means energy density
change due to acceleration is much larger than the one due to pair creation,%
\begin{equation}
Eenv\gg m\gamma S.   \label{weak}
\end{equation}
In this case oscillations equations (\ref{ndot})-(\ref{Edot}) simplify. From
(\ref{rhodot}) and (\ref{pdot}) we have $\dot{\rho}=v\dot{p}$, and using (%
\ref{veleq}) obtain $v=\pm1$. This is the limit when rest mass energy is
much smaller than the kinetic energy, $\gamma\gg1$.

One may therefore use only the first and the last equations from the above
set. Taking time derivative of the Maxwell equation we arrive to a single
second order differential equation 
\begin{equation}
\ddot{E}+\frac{2em^{4}}{\pi^{2}}\left( \frac{E}{E_{c}}\right) \left\vert 
\frac{E}{E_{c}}\right\vert \exp\left( -\pi\left\vert \frac{E_{c}}{E}%
\right\vert \right) =0.   \label{eeq}
\end{equation}
Equation (\ref{eeq}) is integrated numerically to find the length of
oscillations shown in fig. \ref{fig2} for $E<E_{c}$. Notice that condition (%
\ref{weak}) means ultrarelativistic approximation for electrons and
positrons, so that although according to (\ref{ndot}) there is creation of
pairs with rest mass $2m$ for each pair, the corresponding increase of
plasma energy is neglected, as can be seen from (\ref{weak}).

Now we turn to qualitative properties of the system (\ref{ndot})-(\ref{Edot}%
). These nonlinear ordinary differential equations describe certain
dynamical system which can be studied by using methods of qualitative
analysis of dynamical systems. The presence of the two integrals (\ref{rhopn}%
) and (\ref{energy}) allows reduction of the system to two dimensions. It is
useful to work with the variables $v$ and $E$. In these variables we have%
\begin{align}
\frac{d\tilde{v}}{d\tilde{t}} & =\left( 1-\tilde{v}^{2}\right) ^{3/2}\tilde{E%
}, \\
\frac{d\tilde{E}}{d\tilde{t}} & =-\frac{1}{2}\tilde{v}\left( 1-\tilde {v}%
^{2}\right) ^{1/2}\left( \tilde{E}_{0}^{2}-\tilde{E}^{2}\right) -8\pi\alpha%
\frac{\tilde{S}}{\tilde{E}\left( 1-\tilde{v}^{2}\right) ^{1/2}}.
\end{align}
Introducing the new time variable $\tau$%
\begin{equation}
\frac{d\tau}{d\tilde{t}}=\left( 1-\tilde{v}^{2}\right) ^{-1/2}
\end{equation}
we arrive at%
\begin{align}
\frac{d\tilde{v}}{d\tau} & =\left( 1-\tilde{v}^{2}\right) ^{2}\tilde {E},
\label{vdeq} \\
\frac{d\tilde{E}}{d\tau} & =-\frac{1}{2}\tilde{v}\left( 1-\tilde{v}%
^{2}\right) \left( \tilde{E}_{0}^{2}-\tilde{E}^{2}\right) -8\pi\alpha \frac{%
\tilde{S}}{\tilde{E}}.   \label{edeq}
\end{align}
Clearly the phase space is bounded by the two curves $\tilde{v}=\pm1$.
Moreover, physical requirement $\rho\geq0$ leads to existence of two other
bounds $\tilde{E}=\pm\tilde{E}_{0}$. This system has only one singular point
in the physical region, of the type focus at $\tilde{E}=0$ and $\tilde{v}=0$.

\begin{figure}[ptb]
\centering
\includegraphics[width=5in]{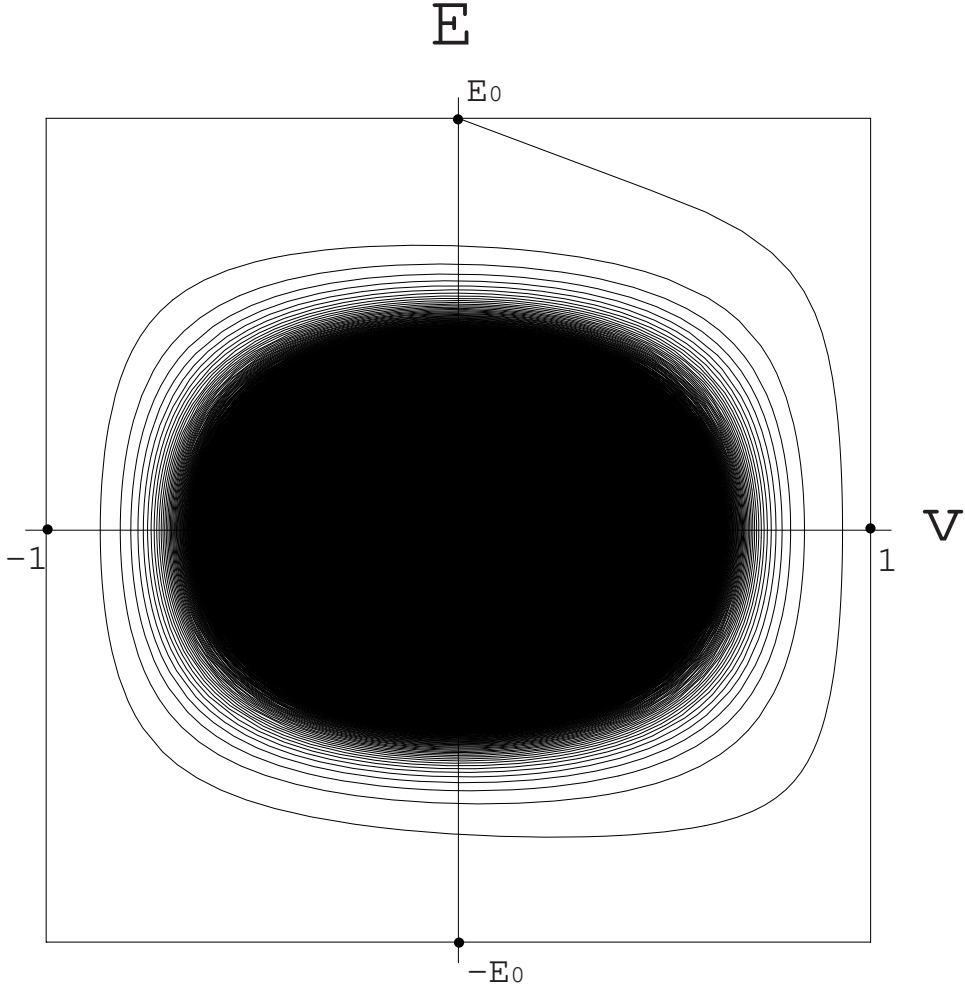}
\caption{Phase portrait of the two-dimensional dynamical system (\protect\ref%
{vdeq}),(\protect\ref{edeq}). Tildes are ommitted. Notice that phase
trajectories are not closed curves and with each cycle they approach the
point with $\tilde{E}=0$ and $\tilde{v}=0$.}
\label{fig5}
\end{figure}

The phase portrait of the dynamical system (\ref{vdeq}),(\ref{edeq}) is
represented at fig. {\ref{fig5}}. 
Thus, every phase trajectory tends asymptotically to the only singular point
at $\tilde{E}=0$ and $\tilde{v}=0$. This means oscillations stop only when
electric field vanishes. At that point clearly 
\begin{equation}
\rho=mn.   \label{rest}
\end{equation}
is valid. i.e. all the energy in the system transforms just to the rest mass
of the pairs.

\begin{figure}[ptb]
\centering
\includegraphics[width=5in]{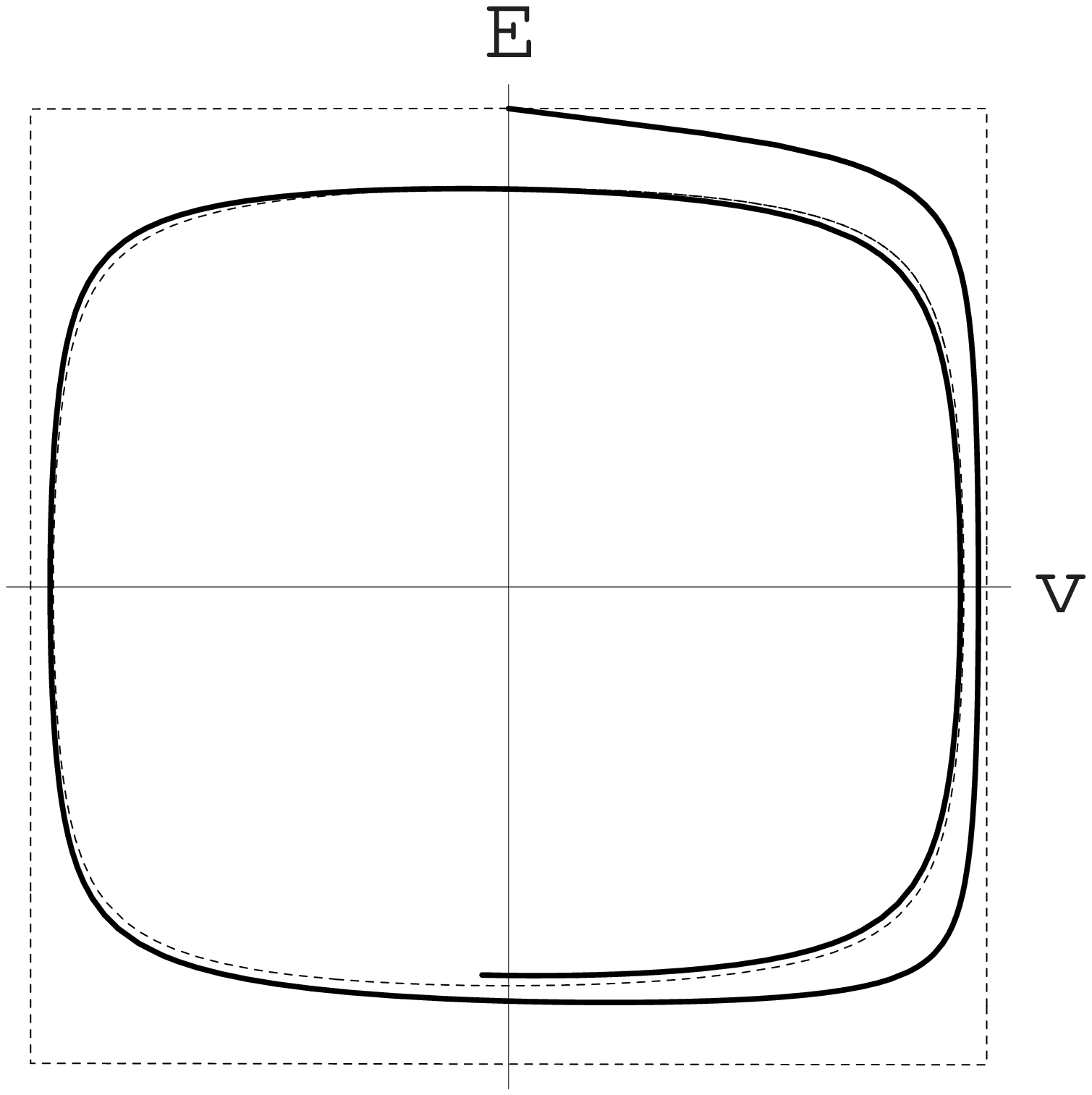}
\caption{Phase trajectory for 1.5 cycles (thick curve) compared with
solutions where the Schwinger pair production is switched off (dashed
curves).}
\label{fig5a}
\end{figure}

In order to illustrate details of the phase trajectories shown at fig. \ref%
{fig5} we plot only 1.5 cycles at fig. \ref{fig5a}. One can see that the
deviation from closed curves shown by dashed curves is maximal when the
field peaks, namely when the pair production rate is maximal. 

The above treatment has been done by considering uniquely back reaction of
the electron-positron pairs on the external uniform electric field. The only
source of damping of the oscillations is pair production, i.e. creation of
mass. As our analysis shows the damping in this case is exponentially weak.
However, since electrons and positrons are strongly accelerated in electric
field the bremsstrahlung radiation may give significant contribution to the
damping of oscillations and further reduce the pair creation rate.
Therefore, the effective rate shown in fig. \ref{fig4} will represent an
upper limit. In order to estimate the effect of bremsstrahlung we recall the
classical formula for the radiation loss in electric field%
\begin{equation}
I=\frac{2}{3}\frac{e^{4}}{m^{2}}E^{2}=\frac{2}{3}\alpha m^{2}\left( \frac{E}{%
E_{c}}\right) ^{2}.
\end{equation}%
Thus the equations (\ref{rhodot}) and (\ref{pdot}), generalized for
bremmstrahlung, are%
\begin{align}
\dot{\rho}& =E\left( env+\frac{m\gamma S}{E}\right) -\frac{2}{3}e^{4}mE^{2},
\label{energy_br} \\
\dot{p}& =enE+mv\gamma S-\frac{2}{3}e^{4}mE^{2}v.  \label{momentum_br}
\end{align}%
while equations (\ref{ndot}) and (\ref{Edot}) remain unchanged. Assuming
that new terms are small, relations (\ref{rhopn}) and (\ref{energy}) are
still approximately satisfied.

Now damping of the oscillations is caused by two terms:%
\begin{equation}
\frac{\tilde{\gamma}}{4\pi ^{2}}\tilde{E}^{2}\exp \left( -\frac{\pi }{\tilde{%
E}}\right) \text{ \ \ \ \ \ and \ \ \ }\frac{2}{3}\alpha \tilde{E}^{2}.
\label{terms}
\end{equation}

\begin{figure}[tbp]
\centering
\includegraphics[width=5in]{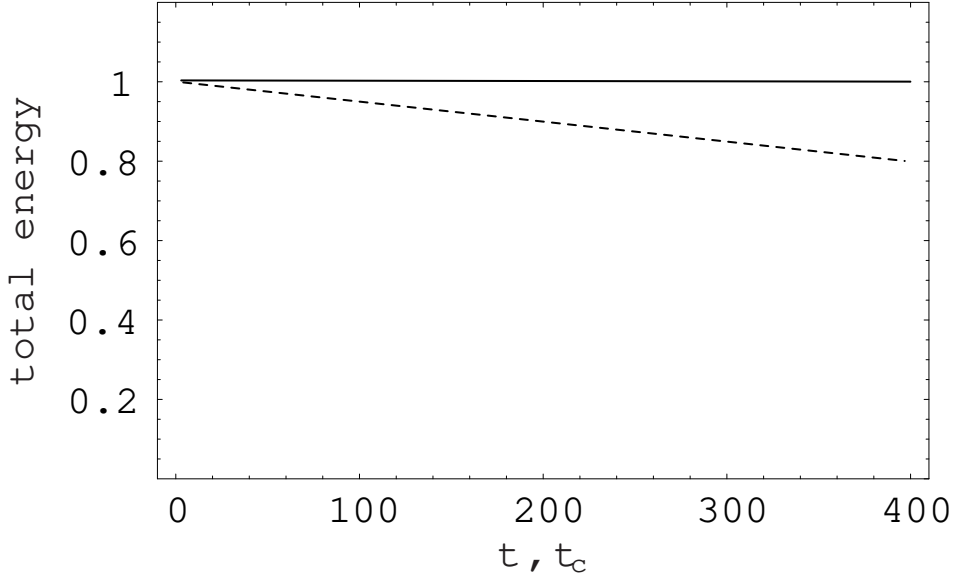}
\caption{Losses of the energy due to classical bremsstrahlung radiation. The
energy density of the system of electrons, positrons and the electric field
normalized to the initial energy density is shown without (solid line) and
with (dashed line) the effect of bremsstrahlung.}
\label{fig6}
\end{figure}
We integrate the modified system of equations, taking into account radiation
loss, starting with $E_{0}=10E_{c}$. We present the results in fig. \ref%
{fig6} where the sum of the energy of electric field and electrons-positrons
pairs normalized to the initial energy is shown as a function of time. The
energy loss reaches 20 percent for 400 Compton times. Thus the effect of
bremmstrahlung is as important as the effect of collisions considered in 
\cite{Astro} for $E>E_{c}$, leading to comparable energy loss for pairs on
the same timescale. For $E<E_{c}$ we expect that the damping due to
bremmstrahlung dominates, but the correct description in this case requires
Vlasov-Boltzmann treatment \cite{ARV}.

The damping of the plasma oscillations due to electron-positron annihilation
into photons has been addressed in \cite{Astro}. There it was found that the
system evolves towards an electron-positron-photon plasma reaching energy
equipartition. Such a system undergoes self-acceleration process following
the work of \cite{RWSX}.

We can therefore reach the following conclusions:

\begin{itemize}
\item It is usually assumed that for $E<E_{c}$ electron-positron pairs,
created by the vacuum polarization process, move as charged particles in
external uniform electric field reaching arbitrary large Lorentz factors. We
demonstrate the existence of plasma oscillations of the elecron-positron
pairs also for $E\lesssim E_{c}$. The corresponding results for $E>E_c$ are well known in the literature. For both cases we determine the maximum Lorentz factors $\gamma _{\max }$
reached by electrons and positrons. The length of oscillations is 10 $\hbar /(mc)$ for $E_{0}=10E_{c}$, and 10$^{7}$ $\hbar /(mc)$ for $E_{0}=0.15E_{c}$. We also study the 
asymptotic behaviour in time, $t\rightarrow\infty$, of the plasma oscillations by the phase portrait technique.

\item For $E>E_{c}$ the vacuum polarization process transforms the
electromagnetic energy of the field mainly in the rest mass of pairs, with
moderate contribution to their kinetic energy:  for $E_{0}=10E_{c}$ we find $\gamma _{\max
}=76$. For $E<E_{c}$ the kinetic energy contribution is maximized with respect to the rest mass of pairs: $\gamma _{\max }=8\times 10^{5}$ for $E_{0}=0.15E_{c}$.

\item In the case of oscillations the effective rate of pair production is
smaller than the rate in uniform electric field a constant in time, and
consequently, the discharge process lasts longer. The half-life of
oscillations is $10^{3}t_{c}$ for $E_{0}=10E_{c}$ and $10^{5}t_{c}$ for $%
E_{0}=0.8E_{c}$. We computed the efficiency of pair production with respect to the one in a uniform constant field. For $E=0.3E_{c}$ the efficiency is reduced to one percent, decreasing further for smaller initial electric field.

\end{itemize}

All these considerations apply to a uniform electric field unbounded in space.
The presence of a boundary or a gradient in electric field would require the use of parial differential equations, in contrast to the ordinary differential equations used here. This topic needs further study. We also estimated the effect of bremsstrahlung for $E>E_{c}$, and found that it represents \ comparable contribution to the damping of the plasma oscillations caused by collisions \cite{Astro}. It is therefore clear, that the effects of oscillations introduces a new and firm upper limit to the rate of pair production which would be further reduced if one takes into account bremsstrahlung, collisions and boundary effects.

We thank the anonymous referee for helpful suggestions on the presentation
of our results.


\begin{thebibliography}{99}
\bibitem{osc} A. Bia\l as and W. Czy\.{z}, \textit{Phys. Rev.} \textbf{D30}
(1984) 2371.

\bibitem{osc1} A. Bia\l as, W. Czy\.{z}, A. Dyrek and W. Florkowski, \textit{%
Nucl. Phys.} \textbf{B296} (1988) 611.

\bibitem{osc2} Y. Kluger, J. M. Eisenberg, B. Svetitsky, F. Cooper and E.
Mottola, \textit{Phys. Rev. Lett.} \textbf{67} (1991) 2427; \textit{Phys.
Rev.} \textbf{D45} (1992) 4659.

\bibitem{QFT} F. Cooper and E. Mottola, P\textit{hys. Rev.} \textbf{D40}
(1989) 456.

\bibitem{ion} T. S. Biro, H. B. Nielsen and J. Knoll, \textit{Nucl. Phys.} 
\textbf{B245} (1984) 449.

\bibitem{ion1} G. Gatoff, A.K. Kerman, T. Matsui, \textit{Phys. Rev.} 
\textbf{D36} (1987) 114.

\bibitem{ion2} F. Cooper, J. M. Eisenberg, Y. Kluger, E. Mottola, and B.
Svetitsky, \textit{Phys. Rev.} \textbf{D48} (1993) 190.

\bibitem{Laser} A. Ringwald, \textit{Phys. Lett.} \textbf{B510} (2001) 107.

\bibitem{Astro} R. Ruffini, L. Vitagliano, S.-S. Xue, \textit{Phys. Lett.} 
\textbf{B559} (2003) 12.

\bibitem{Ringwald} A. Ringwald, hep-ph/0304139.

\bibitem{Laser1} S. S. Bulanov, N. B. Narozhny, V. D. Mur and V. S. Popov, 
\textit{ZhETF} \textbf{129} (2006) 14 [\textit{JETP} \textbf{102} (2006) 9];
Phys.Lett. A330 (2004) 1.

\bibitem{Laser2} D. B. Blaschke, A. V. Prozorkevich, C. D. Roberts, S. M.
Schmidt and S. A. Smolyansky, \textit{Phys. Rev. Lett.} \textbf{96} (2006)
140402.

\bibitem{Gies} H. Gies, J. Jaeckel, A. Ringwald, \textit{Europhys.Lett.} 
\textbf{76} (2006) 794.

\bibitem{Bulanov} S. S. Bulanov, A. M. Fedotov, F. Pegoraro, \textit{%
Phys.Rev.} \textbf{E71} (2005) 016404.

\bibitem{dya2} R. Ruffini, L. Vitagliano, S.-S. Xue, \textit{Phys. Lett.} 
\textbf{B573} (2003) 33.

\bibitem{Tajima} T. Tajima, G. Mourou, \textit{Physical Review ST Accel.
Beams}, \textbf{5} (2002) 031301.

\bibitem{Bulanov2} S.V. Bulanov, T. Esirkepov, T. Tajima, \textit{Physical
Review Letters} \textbf{91} (2003) 085001.

\bibitem{S} F. Sauter, \textit{Z. Phys.} \textbf{69} (1931) 742.

\bibitem{S1} W. Heisenberg, H. Euler, \textit{Z. Phys.} \textbf{98} (1935)
714.

\bibitem{S2} J. Schwinger, \textit{Phys. Rev.} \textbf{82} (1951) 664.

\bibitem{S3} N.B. Narozhnyi, A.I. Nikishov, \textit{Sov. J. Nucl. Phys.} 
\textbf{11} (1970) 596.

\bibitem{book1} W. Greiner, B. M\"{u}ller, and J. Rafelski, \textit{Quantum
Electrodynamics of Strong Fields} \ (Springer-Verlag, Berlin, 1985).

\bibitem{book2} A.A. Grib, S.G. Mamaev, and V.M. Mostepanenko, \textit{%
Vacuum Quantum Effects in Strong External Fields} (Atomizdat, Moscow, 1980).

\bibitem{Berezin} E. Brezin and C. Itzykson, \textit{Phys. Rev.} \textbf{D2}
(1970) 1191.

\bibitem{Popov} V. S. Popov, \textit{JETP Lett.} \textbf{13} (1971) 185; 
\textit{JETP Lett.} \textbf{18} (1973) 255.

\bibitem{Pair} A. I. Nikishov, \textit{ZhETF} \textbf{57} (1969) 1210 [%
\textit{JETP} \textbf{30} (1969) 660].

\bibitem{Pair1} I. B. Khriplovich, \textit{Il Nuovo Cimento} \textbf{B115}
(2000) 761.

\bibitem{ARV} A.G. Aksenov, R. Ruffini, G.V. Vereshchagin, in preparation.

\bibitem{RWSX} R. Ruffini, J. D. Salmonson, J. R. Wilson, and S.-S. Xue, 
\textit{A\&A }\textbf{350} (1999) 334; \textit{A\&A} \textbf{359} (2000) 855.
\end{thebibliography}
\end{document}